\documentclass[fleqn,preprint,showpacs,preprintnumbers,amsmath,amssymb]{revtex4}
\usepackage{graphicx}
\usepackage{dcolumn}
\usepackage{bm}

\begin{document}

\title{Nonlinear ion waves in Fermi-Dirac pair plasmas}
\author{M. Akbari-Moghanjoughi}
\affiliation{ Azarbaijan University of
Tarbiat Moallem, Faculty of Sciences,
Department of Physics, 51745-406, Tabriz, Iran}

\date{\today}
\begin{abstract}
Arbitrary amplitude nonlinear ion waves is investigated in an extremely degenerate electron-positron-ion plasma with relativistic electrons/positrons and dynamic cold ions using Sagdeev pseudo-potential method in framework of quantum hydrodynamics model. The matching criteria of existence for ion solitary as well as periodic nonlinear excitations is studied numerically in terms of the relativistic degeneracy parameter, relative positron concentration and the relativistic normalized propagation-speed range of these waves are obtained. It is remarked that the electron relativistic degeneracy and relative positron concentration in such plasmas have significant effects on nonlinear wave propagations in relativistically degenerate Fermi-Dirac plasma. Our results are in good agreement with the previously reported ones obtained using semiclassical Thomas-Fermi approximation in dense plasmas with non-relativistic and ultra-relativistic electrons and positrons. Current findings can be appropriate for the study of astrophysical superdense compact objects such as white dwarfs.
\end{abstract}


\maketitle

\section{Introduction}

Ion waves are one of the most prominent features of ionized environments with abundant occurrence in astrophysical entities and laboratory plasmas. The characterization of such waves began early in 1961 by work of Vedenov et.al \cite{vedenov} using pseudo-potential method. Another approach which is widely employed to investigate the asymptotic behavior of nonlinear excitations is the so-called reductive perturbation technique \cite{salah, Esfand2, Tiwari1, Tiwari2, Mushtaq}. However, the later approach is suitable only for the study of small-amplitude nonlinear perturbations and is usually applied for treatment of plasma waves in a state very close to thermodynamic equilibrium. In order to achieve satisfactory results applicable to real environments using perturbation method, it is needed to consider the higher-order approximations for dependent plasma variables \cite{Esfand3, akbari1, chatterjee}. In recent years there has been many investigations on arbitrary-amplitude nonlinear ion solitary (ISWs), ion periodic (IPWs) as well as electrostatic solitary waves (ESWs) in different plasma environments \cite{popel, nejoh, mahmood1, mahmood2, mahmood3}. Furthermore, the small amplitude propagation as well as interaction of ISWs has been considered in Ref. \cite{akbari2} in magnetized pair plasmas with the electrons and positrons possessing either non-relativistic or ultra-relativistic degeneracy pressure. It is however the aim of current work to study the propagation of ion waves in dense plasmas with arbitrary degree of relativistic degeneracy in three component electron-positron-ion dense plasmas.

Because of accommodation of variety of nonlinear excitations in pair plasmas with components of equal mass and charge, and also because of their inevitable applications in astrophysics, the electron-positron-ion plasma has attracted outstanding attention in recent years \cite{Shukla, Berezh1, Berezh2, Berezh3, Rizzato}. For instance, electron-positron-ion plasmas exist in active galactic nuclei \cite{miller}, pulsar magnetospheres \cite{michel} and many dense astronomical objects like neutron stars and white dwarfs \cite{Silva, goldr}. It is also believed that the electron-positron-ion plasma may play an important role in the beginning and evolution of Universe \cite{rees}. Pair plasmas have been produced and studied in laboratory \cite{surko1, surko2, greeves, Berezh4}. The electron-positron-ion plasma can also be easily confined in a Tokamak \cite{son1, son2}. Calculations \cite{akbari3} reveal that even in dense electron-positron-ion plasmas ion waves can have significant lifetime to propagate in such environments, despite the high rate of annihilation processes.

On the other hand, in compact dense astrophysical objects \cite{shapiro} and in ordinary metals and semiconductors \cite{haug}, fermion constituents of plasma are under extreme condition called the degeneracy state. In a degenerated plasma the quantum and relativistic effects become important and the much higher degeneracy pressure which arises due to the Pauli-exclusion mechanism should be taken into account \cite{shukla}. The quantum effects become more important, specifically,  when the inter-particle distances become less than the characteristic de Broglie thermal wavelength, $\lambda_D=h/(2\pi m k_B T)^{1/2}$ \cite{Bonitz}. Chandrasekhar \cite{chandra1}, using the Fermi-Dirac statistics for electrons, has given a mathematical criteria under which a white dwarf can be considered as completely degenerate. In the zero-temperature Fermi-gas, used for a completely degenerate plasma, it is assumed that the thermal temperature of plasma species are, although finite, but fall much below the well defined characteristic Fermi-temperature \cite{landau}. Although, the zero-temperature model may seem a simplifying assumption, however, it hold for many dense astrophysical entities such as white dwarfs \cite{chandra2, misner}. It should be noted that to have a more realistic model accounting for the electron/positron temperature, one has to revisit the finite temperature Fermi-Dirac model which can be our future work.

It has been shown that, the electron degeneracy in massive white dwarfs which opposes the catastrophic gravitational inward pressure may give rise to softened relativistic degeneracy pressure leading to the ultimate collapse of massive star in some conditions \cite{chandra2}. There are some studies, using quantum hydrodynamics, which deal with astrophysical models considering only the non-relativistic electron degeneracy case \cite{sabry, misra1, misra2, abdelsalam}. Whereas, in a typical white dwarf with the electron density as high as $10^{28}cm^{-3}$ ($\rho\simeq 10^7(gr/cm^3)$), the degeneracy pressure is extreme and both quantum and relativistic considerations have to be invoked. Application of semiclassical Thomas-Fermi approximation indicates that the dynamics of nonlinear ion excitations in the electron-positron-ion plasma can be quite different when the electrons/positrons are under ultra-relativistic degeneracy pressure from that under normal degeneracy pressure \cite{akbari4}. Previously reported investigations, based on the quantum hydrodynamics model \cite{Marklund, Shaikh, Brodin, Gardner}, consider normal degeneracy case encountered in ordinary metals and semiconductors. In this study, we investigate a degenerated plasma for a whole range of the relativity parameter (degree of electron degeneracy) which can extend the applications of quantum hydrodynamics model used in mentioned references to extreme conditions found in compact astrophysical objects. We also use the Fermi-Dirac statistics for electrons and positrons instead of the simplified Thomas-Fermi approximation. The presentation of the article is as follows. The basic normalized plasma equations are introduced in section \ref{equations}. Nonlinear arbitrary-amplitude solutions are derived in section \ref{Sagdeev}. Numerical analysis and discussions is given in section \ref{discussion} and final remarks are presented in section \ref{conclusion}.

\section{Quantum Hydrodynamics Model}\label{equations}

In this section we use quantum hydrodynamics (QHD) model to study the nonlinear dynamics in a superdense completely degenerate electron-positron-ion plasma. The electrons/positrons are considered relativistic and follow the Fermi-Dirac statistics and the positively charged cold-ions are dynamic. In this model the rate of electron(positron)-ion collisions are limited due to Pauli-blocking process and the pair-annihilations can be ignored \cite{akbari3}. We therefore write the conventional QHD fluid equations in dimensional form as
\begin{equation}\label{dim}
\begin{array}{l}
\frac{{\partial {n_i}}}{{\partial t}} + \frac{{\partial {n_i}{v_i}}}{{\partial x}} = 0, \\
\frac{{\partial {n_j}}}{{\partial t}} + \frac{{\partial {n_j}{v_j}}}{{\partial x}} = 0, \\
\frac{{\partial {v_i}}}{{\partial t}} + {v_i}\frac{{\partial {v_i}}}{{\partial x}} =  - \frac{e}{{{m_i}}}\frac{{\partial \phi }}{{\partial x}} - \frac{{\gamma {k_B}{T_i}}}{{{m_i n_{i0}}}}\left(\frac{n_i}{n_{i0}}\right)^{\gamma  - 2}\frac{{\partial {n_i}}}{{\partial x}}, \\
\frac{{{m_j}}}{{{m_i}}}\left( {\frac{{\partial {v_j}}}{{\partial t}} + {v_j}\frac{{\partial {v_j}}}{{\partial x}}} \right) = - \frac{e s_j}{{{m_i}}}\frac{{\partial \phi }}{{\partial x}} - \frac{1}{{{m_i}{n_j}}}\frac{{\partial {P_j}}}{{\partial x}} + {\frac{{{m_j}}}{{{m_i}}}}\left(\frac{{{\hbar ^2}}}{{2{m_j^{2}}}}\right)\frac{\partial }{{\partial x}}\left( {\frac{1}{{\sqrt {{n_j}} }}\frac{{{\partial ^2}\sqrt {{n_j}} }}{{\partial {x^2}}}} \right), \\ \frac{{{\partial ^2}\phi }}{{\partial {x^2}}} = - 4\pi e\left( {\sum\limits_j {{s_j}{n_j}}  + {n_i}} \right), \\
\end{array}
\end{equation}
where, $n_{i,j}$, $m_{i,j}$ and $v_{i,j}$ denote the ions and $j$-species number-density, mass and velocities with $i$ for ions and $j=\{e,p\}$ with $s_j=\{-1,+1\}$, for electrons and positrons, respectively. Also $\gamma$, $T_i$ and $\hbar$, respectively, present the adiabatic constant, ion-temperature and the scaled Plank-constant. The normalized set of QHD equations follow from the following scaling
\begin{equation}\label{T}
x \to \frac{{{c_{sr}}}}{{{\omega _{pi}}}}\bar x,\hspace{3mm}t \to \frac{{\bar t}}{{{\omega _{pi}}}},\hspace{3mm}n_{e,p,i} \to \bar n_{e,p,i}{n_{e0}},\hspace{3mm}v_{i,e,p} \to \bar v_{i,e,p}{c_{sr}},\hspace{3mm}\phi  \to \bar \phi \frac{{{m_e}{c^2}}}{e}.
\end{equation}
where, $c$, ${\omega _{pi}} = \sqrt {4\pi {e^2}{n_{e0}}/{m_i}}$ and ${c_{sr}} = \sqrt {{m_e} {c^2}/{m_i }}$ are the vacuum light speed and the characteristic plasma frequency and sound-speed, respectively, and the parameter $n_{e0}$ denotes the electron equilibrium density. The bar symbols denote the dimensionless quantities which are omitted in forthcoming algebra for simplicity. In a relativistic completely degenerate Fermi-Dirac plasma, the electron/positron relativistic degeneracy pressure is expressed in the following form \cite{chandra1}
\begin{equation}\label{p}
{P_j} = \frac{{\pi m_j ^4{c^5}}}{{3{h^3}}}\left\{ {{R_j}\left( {2{R_j^{2}} - 3} \right)\sqrt {1 + {R_j^{2}}}  + 3\text{sinh}^{-1}{R_j}} \right\},
\end{equation}
in which $R_j=(n_j/n_0)^{1/3}$ (${n_0} = \frac{{8\pi m_\alpha^3{c^3}}}{{3{h^3}}}\simeq 5.9 \times 10^{29} cm^{-3}$) is the ratio of electron/positron Fermi relativistic momentum to $m_e c$. In the limits of very small and very large values of $R_j$ we obtain
\begin{equation}\label{limits}
P_j = \left\{ {\begin{array}{*{20}{c}}
{\frac{1}{{20}}{{\left( {\frac{3}{\pi }} \right)}^{\frac{2}{3}}}\frac{{{h^2}}}{{{m_j}}}n_j^{\frac{5}{3}}\hspace{10mm}(R_j \to 0)}  \\
{\frac{1}{{8}}{{\left( {\frac{3}{\pi }} \right)}^{\frac{1}{3}}}hcn_j^{\frac{4}{3}}\hspace{10mm}(R_j \to \infty )}  \\
\end{array}} \right\}.
\end{equation}
Thus, the normalized QHD set of plasma equations including the electron/poitron tunneling effect by making use of the fact that $(1/{n_j}){\partial _x}{P_j} = {\partial _x}\sqrt {1 + R_j^2}$, $R_j=R_0 (\alpha_j n_j)^{1/3}$ and $\alpha_j=\{1, \alpha\}$ for equilibrium number-densities of electrons and positrons, respectively, with $\alpha=n_{p0}/n_{e0}$ being the relative positron to electron number-density, can be written in the following form
\begin{equation}\label{normal}
\begin{array}{l}
\frac{{\partial {n_i}}}{{\partial t}} + \frac{{\partial {n_i}{v_i}}}{{\partial x}} = 0, \\
\frac{{\partial {n_j}}}{{\partial t}} + \frac{{\partial {n_j}{v_j}}}{{\partial x}} = 0, \\
\frac{{\partial {v_i}}}{{\partial t}} + {v_i}\frac{{\partial {v_i}}}{{\partial x}} =  - \frac{{\partial \phi }}{{\partial x}} - \left( {\frac{{\gamma {k_B}{T_i}}}{{n_{i0}{m_e}{c^2}}}} \right)\left(\frac{n_i}{n_{i0}}\right)^{\gamma  - 2}\frac{{\partial {n_i}}}{{\partial x}}, \\
\frac{{{m_j}}}{{{m_i}}}\left( {\frac{{\partial {v_j}}}{{\partial t}} + {v_j}\frac{{\partial {v_j}}}{{\partial x}}} \right) = -s_j \frac{{\partial \phi }}{{\partial x}} - \frac{{\partial}}{{\partial x}} \left[ {1 + {R_0^{2}\alpha_j^{2/3} {n_j}^{2/3}}} \right]^{1/2} + {\frac{{{m_j}}}{{{m_i}}}}\left( \frac{{{H_r^{2}}}}{2}\right)\frac{\partial }{{\partial x}}\left( {\frac{1}{{\sqrt {{n_e}} }}\frac{{{\partial ^2}\sqrt {{n_e}} }}{{\partial {x^2}}}} \right), \\
\frac{{{\partial ^2}\phi }}{{\partial {x^2}}} = - 4\pi e\left( {\sum\limits_j {{s_j}{n_j}}  + {n_i}} \right), \\
\end{array}
\end{equation}
where, we have introduced the normalized degeneracy parameter, $R_0=(n_{e0}/n_0)^{1/3}$, as a measure of the relativistic degeneracy effects i.e. so called the normalized relativity parameter. Therefore, the parameter $R_0^{3}$ denotes the relative electron number density with respect to the base value $n_0$.

Furthermore, the parameter, $(m_e/m_i)H_r^{2}/2=(m_e/m_i)\hbar^2 \omega_{pi}^2/(m_e c^2)^2$, is the quantum diffraction coefficient which is related to the relativity parameter via; $(m_e/m_i)H_r^{2}/2\simeq 4.6\times 10^{-10} R_0^{3}$. It should be noted that the parameter $R_0$ is related also to the mass-density (of white dwarf, for instance) through the electron number-density with relation $\rho\simeq 2m_p n_{e0}(1-\alpha)$ or $\rho(gr/cm^{3})=(1-\alpha)\rho_0 R_0^{3}$ with $\rho_0(gr/cm^{3})\simeq 1.97\times 10^6$, where, $m_p$ is the proton mass. The density $\rho_0$ is in the range of a mass-density of a typical white dwarf and a contribution of $(1-\alpha)$ has been included in mass-density definition to account for the electron-positron pair production/anihilation. The density of typical white dwarfs can be in the range $10^5<\rho(gr/cm^{3})<10^{8}$, which, neglecting electron-positron pair production/anihilation ($\alpha\ll 1$), results in values of $0.37<R_0<8$ for the relativity parameter of a typical white dwarf. S. Chandrasekhar, in ground of Fermi-Dirac statistics, has proven that for a white dwarf with a mass density $\rho$, the electron degeneracy pressure turns from $P_e\propto \rho^{5/3}$ dependence for normal degeneracy ($R_0 \ll 1$) to $P_e\propto \rho^{4/3}$ dependence for ultra-relativistic electron degeneracy ($R_0 \gg 1$) \cite{chandra2}. In this calculation we show that this behavior has also a profound effect on the propagation and stability of nonlinear ion-acoustic excitations in a white dwarf. Now, neglecting the terms containing the small mass-ratio, $m_e/m_i$ and assuming that $k_B T_i \ll m_e c^2$, we arrive at the following simplified set of basic normalized equations
\begin{equation}\label{comp}
\begin{array}{l}
\frac{{\partial {n_i}}}{{\partial t}} + \frac{{\partial {n_i}{v_i}}}{{\partial x}} = 0, \\
\frac{{\partial {n_j}}}{{\partial t}} + \frac{{\partial {n_j}{v_j}}}{{\partial x}} = 0, \\
\frac{{\partial {v_i}}}{{\partial t}} + {v_i}\frac{{\partial {v_i}}}{{\partial x}} =  - \frac{{\partial \phi }}{{\partial x}}, \\
-s_j \frac{{\partial \phi }}{{\partial x}} - \frac{{\partial}}{{\partial x}}\left[ {1 + {R_0^{2} \alpha_j^{2/3}{n_j}^{2/3}}}\right]^{1/2} =0, \\
\frac{{{\partial ^2}\phi }}{{\partial {x^2}}} = - \sum\limits_j {{s_j}{n_j}}  - {n_i}. \\
\end{array}
\end{equation}
\textbf{In order to get the steady state solution to the hydrodynamics normalized set of fluid equations, Eqs. (\ref{comp}), we change to a stationary frame of soliton by changing the coordinate to $\xi=x-Mt$, with, $M$ being the constant normalized speed of wave train relative to the quantity, $c_{sr}$ (not be confused with the Much-number parameter as pointed out in Ref. \cite{dubinov}). Therefore, after trivial integration with the appropriate boundary requirements; $\mathop {\lim }\limits_{{v_{i,e,p}} \to 0} {n_{i,e,p}} = \{1-\alpha,1,\alpha\}$ and $\mathop {\lim }\limits_{{v_{i,e,p}} \to 0} \phi  = 0$, we obtain}
\begin{equation}\label{red}
\begin{array}{l}
M\left[ {{n_i} - (1 - \alpha )} \right] = {n_i}{v_i}, \\
M{v_i} - \frac{{v_i^2}}{2} = \phi , \\
{s_j}\phi  = - \sqrt {1 + R_0^2\alpha _j^{2/3}{n_j}^{2/3}}.
\end{array}
\end{equation}
Solving Eqs. (\ref{red}) for the ion, electron and positron number-densities in terms of electrostatic potential leads to the following energy-density relations
\begin{equation}\label{phis}
\begin{array}{l}
{n_i} = (1 - \alpha ){\left( {1 - \frac{{2\phi }}{{{M^2}}}} \right)^{ - 1/2}}, \\
{n_e} = R_0^{ - 3}{\left[ {R_0^2 + \phi \left( {2\sqrt {1 + R_0^2}  + \phi } \right)} \right]^{3/2}} \\
{n_p} = {\alpha ^{ - 1}}R_0^{ - 3}{\left[ {R_0^2{\alpha ^{4/3}} - \phi (2\sqrt {1 + R_0^2{\alpha ^{4/3}}}  - \phi )} \right]^{3/2}}. \\
\end{array}
\end{equation}
These equations are completely different from the energy relations frequently used in semiclassical Thomas-Fermi model \cite{akbari2, akbari3, akbari4, abdelsalam}, as it should be. Note that, unlike for Thomas-Fermi model for plasma, in Eqs. (\ref{phis}) there is no explicit contribution for electron/positron (Fermi) temperatures. In fact there is a relation between the Fermi-temperatures and the number-densities of quantum species i.e. ${T_{Fj}} = {E_{Fj}}/{k_B} = ({{{\hbar ^2}}}/({{2{m_j}{k_B}}})){(3{\pi ^2}{n_{j,0}})^{3/2}}$. In the following section we will find the appropriate pseudo-potential, based on relations Eqs. (\ref{phis}) and Poisson's condition, to evaluate the propagation of nonlinear ion-acoustic waves in the plasma described by Eq. (\ref{comp}).

\section{Pseudo-Potential Approach}\label{Sagdeev}

Proceeding with the pseudo-potential approach, we aim at obtaining the appropriate energy integral based on the zero-temperature Fermi-Dirac statistics for electrons and positrons. The solution to this integral yields a description for the propagation of arbitrary-amplitude ISWs as well as IPWs periodic structures in a relativistically degenerate electron-positron-ion plasma. Using the density relations given in Eq. (\ref{phis}) and substituting into Poisson's relation, we obtain
\begin{equation}\label{int}
\begin{array}{l}
\frac{{{d^2}\phi }}{{d{\xi ^2}}} = {\left\{ {R_0^{ - 3}{{\left[ {R_0^2 + \phi \left( {2\sqrt {1 + R_0^2}  + \phi } \right)} \right]}^{3/2}}} \right.}  \\ \left. {{ - {\alpha ^{ - 1}}R_0^{ - 3}{{\left[ {R_0^2{\alpha ^{4/3}} - \phi \left( {2\sqrt {1 + {\alpha ^{4/3}}R_0^2}  - \phi } \right)} \right]}^{3/2}}} - \frac{{1 - \alpha }}{{\sqrt {1 - \frac{{2\phi }}{{{M^2}}}} }}} \right\} \\
\end{array}
\end{equation}
Now, multiplying Eq. (\ref{int}) by $\frac{{d\phi }}{{d\xi }}$ and integrating once leads to the desired energy integral in terms of pseudo-coordinates ($\phi,\xi$)
\begin{equation}\label{energy1}
\begin{array}{l}
\frac{1}{2}{\left( {\frac{{d\phi }}{{d\xi }}} \right)^2} - \int_0^\phi  {\left\{ {R_0^{ - 3}{{\left[ {R_0^2 + \phi \left( {2\sqrt {1 + R_0^2}  + \phi } \right)} \right]}^{3/2}}} \right.}  \\
\left. {{ - {\alpha ^{ - 1}}R_0^{ - 3}{{\left[ {R_0^2{\alpha ^{4/3}} - \phi \left( {2\sqrt {1 + {\alpha ^{4/3}}R_0^2}  - \phi } \right)} \right]}^{3/2}}} - \frac{{1 - \alpha }}{{\sqrt {1 - \frac{{2\phi }}{{{M^2}}}} }}} \right\}{\rm{d}}\phi = 0  \\
\end{array}
\end{equation}
or, the well-known energy integral reads as
\begin{equation}\label{energy}
\frac{1}{2}{\left( {\frac{{d\phi }}{{d\xi }}} \right)^2} + U(\phi ) = 0,
\end{equation}
with the long expression for the pseudo-potential $U(\phi)$ given as
\begin{equation}\label{S1}
\begin{array}{l}
U(\phi ) =  \frac{1}{8}\left\{ {8M(\alpha  - 1)\sqrt {{M^2} - 2\phi } } \right. + 3{R_{0}^{ - 3}}\ln M\left({{M^2} - 2\phi } \right) - 3{R_{0}^{ - 3}}{\alpha ^{ - 1}}\ln M\left({{M^2} - 2\phi } \right) \\
- {R_{0}^{ - 3}}\left( {\sqrt {1 + {R_{0}^2}}  + \phi } \right)\left[ {2\phi \left( {2\sqrt {1 + {R_{0}^2}}  + \phi } \right) + 2{R_{0}^2}} \right]\sqrt {{R_{0}^2} - 3 + \phi \left( {2\sqrt {1 + {R_{0}^2}}  + \phi } \right)}  \\
- {R_{0}^{ - 3}}{\alpha ^{ - 1}}\left( {\sqrt {1 + {R_{0}^2}{\alpha ^{4/3}}}  - \phi } \right)\sqrt {{R_{0}^2}{\alpha ^{4/3}} + \phi \left( {\phi  - 2\sqrt {1 + {R_{0}^2}{\alpha ^{4/3}}} } \right)}  \times  \\
\times \left[ {2\phi \left( {\phi  - 2\sqrt {1 + {R_{0}^2}{\alpha ^{4/3}}} } \right) - 3 + 2{R_{0}^2}{\alpha ^{4/3}}} \right]  \\
+ {R_{0}^{ - 3}}{\alpha ^{ - 1}}\left[ {8{M^2}{R_{0}^3}\alpha (1 - \alpha ) - 3R_{0}\left( {\sqrt {1 + {R_{0}^2}} \alpha  + {\alpha ^{2/3}}\sqrt {1 + {R_{0}^2}{\alpha ^{4/3}}} } \right)} \right. \\
+ 2{R_{0}^3}\alpha \left( {\sqrt {1 + {R_{0}^2}}  + \alpha \sqrt {1 + {R_{0}^2}{\alpha ^{4/3}}} } \right) - 3(\alpha  - 1)\ln {M^3} \\
\left. { + 3\alpha \ln 2{M^4}\left( {R_{0} + \sqrt {1 + {R_{0}^2}} } \right) - 3\ln 2{M^4}\left( {R_{0}{\alpha ^{2/3}} - \sqrt {1 + {R_{0}^2}{\alpha ^{4/3}}} } \right)} \right] \\
- 3{R_{0}^{ - 3}}\ln \left[ {2{M^2}({M^2} - 2\phi )\left( {\sqrt {1 + {R_{0}^2}}  + \phi  + \sqrt {{R_{0}^2} + 2\sqrt {1 + {R_{0}^2}} \phi  + {\phi ^2}} } \right)} \right] + 3{R_{0}^{ - 3}} {\alpha ^{ - 1}} \times\\
\left. {\times\ln \left[ {2{M^2}({M^2} - 2\phi )\left( {\sqrt {{R_{0}^2}{\alpha ^{4/3}} - 2\sqrt {1 + {R_{0}^2}{\alpha ^{4/3}}} \phi  + {\phi ^2}}  + \phi  - \sqrt {1 + {R_{0}^2}{\alpha ^{4/3}}} } \right)} \right]} \right\}. \\
\end{array}
\end{equation}
The pseudo-potential given $U(\phi)$ and its first derivative ${\partial _\phi }U(\phi )$ both vanish at $\phi=0$, as expected. From density relations given in Eqs. (\ref{phis}), we may derive the possible maximum/minimum values of the electrostatic potential $\phi$ in which the pseudo-particle of as unit mass is confined, these maximum/minimum values are
\begin{equation}\label{phim1}
\begin{array}{l}
{\phi _{m1}} = \frac{{{M^2}}}{2}\\
{\phi _{m2}} =  \pm 1 + \sqrt {1 + R_0^2}  \\
{\phi _{m3}} =  \pm 1 + \sqrt {1 + R_0^2{\alpha ^{2/3}}}.\\
\end{array}
\end{equation}
The possibility of solitary waves then relies on the following boundary conditions to met, simultaneously
\begin{equation}\label{conditions}
{\left. {U(\phi)} \right|_{\phi = 0}} = {\left. {\frac{{dU(\phi)}}{{d\phi}}} \right|_{\phi = 0}} = 0,{\left. {\frac{{{d^2}U(\phi)}}{{d{\phi^2}}}} \right|_{\phi = 0}} < 0.
\end{equation}
For the existence of the ISWs, it is further required that for at least one either maximum or minimum nonzero $\phi$-value, we have $U(\phi_{m})=0$, so that for every value of $\phi$ in the range ${\phi _m} > \phi  > 0$ or ${\phi _m} < \phi  < 0$, $U(\phi)$ is negative. In such a case, there will be a potential minimum, which describes the propagation of an ion-acoustic solitary excitation.

Also, for IPWs excitations to propagate it is required that
\begin{equation}\label{conditions2}
{\left. {U(\phi)} \right|_{\phi = 0}} = {\left. {\frac{{dU(\phi)}}{{d\phi}}} \right|_{\phi = 0}} = 0,{\left. {\frac{{{d^2}U(\phi)}}{{d{\phi^2}}}} \right|_{\phi = 0}} > 0,
\end{equation}
On the other hand, it is also required that for at least one either maximum or minimum nonzero $\phi$-value, we have $U(\phi_{m})=0$, so that for every value of $\phi$ in the range ${\phi _m} > \phi  > 0$ or ${\phi _m} < \phi  < 0$, $U(\phi)$ is positive. In such a condition we will have potential maximum which describes the propagation of an ion-acoustic periodic structure.

It should be realized already that the derivation of an analytic expression for the matching critical normalized propagation-speed, $M$, form the second derivative of the pseudo-potential given by Eq. \ref{S1} is not possible, hence, we will indulge for the numerical analysis which is presented in the next section.
However, some information may be obtained via the small-amplitude limit. Expanding the potential $U(\phi)$ near $\phi =0$, we obtain
\begin{equation}
U(\phi )=\frac{U''_{0}}{2}\phi^{2}+\frac{U'''_{0}}{6}\phi^{3},
\end{equation}
where, $U''_{0}=U''(\phi=0 )$ and $U'''_{0}=U'''(\phi=0)$. Inserting into Eq. (\ref{energy}) and integrating, we obtain (provided that $U''_{0}<0$) a solitary solution of the form (see \cite{Bertho})
\begin{equation}
\phi(\xi)=-3\frac{U''_{0}}{U'''_{0}}\frac{1}{\cosh^{2}(\frac{1}{2}\sqrt{-U''_{0}}\xi)}
\end{equation}
This pulse profile is equivalent to the stationary soliton solution of the Korteweg-de Vries (KdV) equation, obtained by use of the
reductive perturbation method \cite{Treum}. It is important to notice that the soliton width $\Lambda=2/\sqrt{-U''_{0}}$ and
amplitude $\phi_{0}= -3U''_{0}/U'''_{0}$ satisfy $\phi_{0}\Lambda^{2}=12/U'''_{0} = cst.$, as expected from the KdV theory.

\section{Discussion}\label{discussion}

Figure 1 shows the regions of existence of ISWs/IPWs in yellow/blue (right/left region) for different values of relative positron concentration, $\alpha$, in $M$-$R_0$ plane. It is apparent that the allowed matching normalized propagation-speed range, $M$, is significantly affected by the relativity parameter, $R_0$ as well as the fractional positron concentration, $\alpha$. It is observed that the IPWs posses lower matching normalized propagation-speed value, $M$, compared to ISWs for all values of $R_0$ and the relative positron density, $\alpha$. It is also remarked that, as the relativity parameter increases, the matching normalized propagation-speed range, $M$, for both nonlinear excitation types increase and the allowed matching normalized propagation-speed range, $M$, tends to higher values, while, the effect of the increase in relative positron density on the IPWs is to shift the matching normalized propagation-speed range, $M$, to lower values for all relativity parameter values, while this change on the ISWs is completely different. The similar plots have been presented in Ref. \cite{akbari5} for the relativistically degenerate electron-ion plasma. The increase in the value of small-$\alpha$ is to enhance the matching normalized propagation-speed values, while it is to decrease the values of matching normalized propagation-speed for large-$\alpha$ values. The plots shown in Fig. 2 confirm that there is a maximum matching normalized propagation-speed value for ISWs in $M$-$\alpha$ plane shown for two extreme low- and high-relativistic degeneracy values, namely, $R_0=1$ and $R_0=50$. This maximum matching normalized propagation-speed value is $M_m\simeq 3.75$ for $R_0=1$ while it is diminished to $M_m\simeq 2$ for $R_0=50$ which correspond to ultra-relativistic degeneracy case. Note that, the maximum values found here roughly agree with the corresponding values reported for Thomas-Fermi electron-positron-ion plasmas in non-relativistic and ultra-relativistic degeneracy limits (see Fig. 1(a) in Ref. \cite{akbari3}).

It is further noticed from Fig. 2 that the lower part of plots (smaller $\alpha$-value) is affected more by change in the relativistic factor $R_0$. It should be mentioned that, as the relativity parameter, $R_0$, increases the quantum effects (via the parameter $H_r$) become dominant, hence, the quantum force term in (Eq. (\ref{normal})) can never be ignored for very large value of $R_0$. For instance, for the very large value of $R_0=10^3$ (which may be realized for a collapsing white dwarfs or neutron stars) we get a contribution of order $(m_e/m_i)H_r^{2}/2\simeq 0.46$ for the quantum correction coefficient. Therefore, while Fig. 2(a) qualitatively explains the effect of relativistic degeneracy effects on propagation of nonlinear waves in Fermi-Dirac pair-plasma, a correction must be made for the very highe values of $R_0$. However, it has been shown that the quantum diffraction parameter does not modify the wave amplitude (which increases with increase in the value of relative positron concentration, $\alpha$) and makes only corrections to the width of the nonlinear wave \cite{akbari3}. This leaves only a minor correction for quantum correction which effects all values of positron concentration equally, and therefore does not alter our current qualitative approach.

\textbf{The profiles of pseudo-potentials and their variation with respect to the relativity parameter, relative positron concentration and matching normalized propagation-speed range, $M$, are shown in Fig. 3 for both ISWs and IPWs. As it is clearly remarked, only compressive ISWs propagate in this plasma. It is observed from Figs. 3(a) and 3(b) that for ISWs/IPWs the increase in the relative positron concentration decreases/increases the potential depth and width for other fixed parameters. The same rule applies also to Figs. 3(c) and 3(d), which show the variations of potential width/depth (wave amplitude/width) for ISWs and IPWs with respect to different relative matching-speed values, $M$.} 

\textbf{Figure 4 plots the variations of electrostatic potential (density soliton) profiles with respect to the fractional positron concentration, $\alpha$, and the relativistic degeneracy parameter, $R_0$. From Fig. 4(a) it is clearly observed that, the increase in the values of relative positron concentration, $\alpha$, leads to increase/decrease of amplitude/width of solitary nonlinear structures. On the other hand, from Fig. 4(b) it is revealed that, the increase in the values of relativistic degeneracy parameter, $\alpha$, leads to decrease/increases of amplitude/width of solitary nonlinear structures. It is therefore confirmed that the narrower ISWs the taller it is, which is in accord with the nonlinear theory of soliton in agreement with the relation $\phi_0^{2}\Lambda=cst.$, where, $\phi_0$ and $\Lambda$ denote the amplitude and the width of nonlinear waves, respectively.}

Furthermore, Fig. 5 shows the stability regions for nonlinear ISWs and IPWs for two extreme cases of mass density. The lower $R_0=1$ value corresponds to the mass-density of a typical white-dwarf, while, the higher value of $R_0=50$ which results in a mass-density much above that of a white dwarf, is given to show the effect of higher relativistic degeneracy effects on nonlinear wave dynamics, that is, to consider the ultra-relativistic limit of Eq. (\ref{p}) ($R_0\rightarrow \infty$). It is observed that the stability region for $\rho\simeq 1.97\times 10^{6} (gr/cm^{3})$ lies entirely in subsonic, $v_g<c_{sr}$, limit, while for $\rho\simeq 2.5\times 10^{11} (gr/cm^{3})$ a large portion of ISWs region extends to supper-sonic region, $v_g>c_{sr}$. It should be distinguished that, the speed, $c_{sr}$, is a relativistic speed, despite the name of sound speed and is much higher ($c_{sr}\simeq 8\times 10^{8} (cm/s)$ comparable to the Fermi-speed of electrons in typical solids) compared to that of the ordinary plasmas. It is also remarked that, the inclusion of pair modifies greatly the high positron-concentration portion of the stability regions  (e.g. compare Figs. 2 and 4), as expected. Note that, in calculations for the mass density, $\rho\simeq 2m_p n_{e0}(1-\alpha)$, there is a $\alpha$-dependent term which accounts for the extra electrons produced/annihilated in the plasma. Since, the current analysis is applicable to the wide range of the relativity parameter i.e. the wide plasma mass-density range; it can also be applied to low mass-density degenerate plasma produced in laboratory.
\section{Conclusions}\label{conclusion}

The Sagdeev pseudo-potential approach was used to investigate the propagation of nonlinear ion waves in a relativistically degenerate pair plasma in the framework of quantum hydrodynamics model. The matching criteria of existence of such solitary excitations in terms of the relativistic degeneracy parameter, relative positron density were numerically investigated and the allowed matching normalized propagation-speed range, $M$, for propagation of such waves was evaluated numerically. It was shown that the electron relativistic degeneracy as well as the relative positron concentration has outstanding effects on nonlinear wave dynamics of relativistically degenerate plasmas such as that found in white dwarfs and the cores of massive planets. Current findings are completely consistent with the previously reported similar plasma cases treated with semiclassical assumptions for non-relativistic and ultra-relativistic electrons/positrons. This study can be appropriate for applications in inertial confinement fusion laboratory research, as well as the study of astrophysical dense objects.

\newpage

\textbf{FIGURE CAPTIONS}

\bigskip

Figure-1

\bigskip

(Color online) Variation of the stability regions of arbitrary-amplitude ISWs/IPWs is shown in $M$-$R_0$ plane for a relativistically degenerate electron-positron-ion plasma with respect to different relative positron concentrations, $\alpha$. The region to the right (yellow) correspond to the stability of ISWs and to the left (blue) region to IPWs.

\bigskip

Figure-2

\bigskip

(Color online) The variation of Sagdeev pseudo-potential profile with respect to different relativistic degeneracy parameters, $R_0$, relative positron concentrations $\alpha$ and the matching normalized propagation-speed range, $M$, for both ISWs (left column in figure) and IPWs (right column in figure).

\bigskip

Figure-3

\bigskip

(Color online) Variation of the stability regions of arbitrary-amplitude ISWs/IPWs is shown in $M$-$\alpha$ plane for a relativistically degenerate electron-positron-ion plasma for two extreme relativistic cases with respect to relativistic degeneracy parameter, $R_0$. The region to the right (yellow) correspond to the stability of ISWs and to the left (blue) region to IPWs.

\bigskip

Figure-4

\bigskip

(Color online) (a) The variation of arbitrary-amplitude ion solitary wave profile is shown with respect to different relative positron concentration, $\alpha$ and other fixed plasma parameters. (a) The variation of arbitrary-amplitude ion solitary wave profile is depicted with respect to different relativistic degeneracy parameter, $R_0$ and other fixed plasma parameters. The increase in thickness of curves in each plot represents the increases in the varied plasma parameter.

\bigskip

Figure-5

\bigskip

(Color online) Variation of the stability regions of arbitrary-amplitude ISWs/IPWs is shown in $M$-$\alpha$ plane for a relativistically degenerate electron-positron-ion plasma for two extreme white dwarf mass-densities. The parameter $\rho$ is the corresponding mass-density of a typical white dwarf. The region to the right (yellow) correspond to the stability of ISWs and to the left (blue) region to IPWs.

\newpage

\begin{figure}
\resizebox{1\columnwidth}{!}{\includegraphics{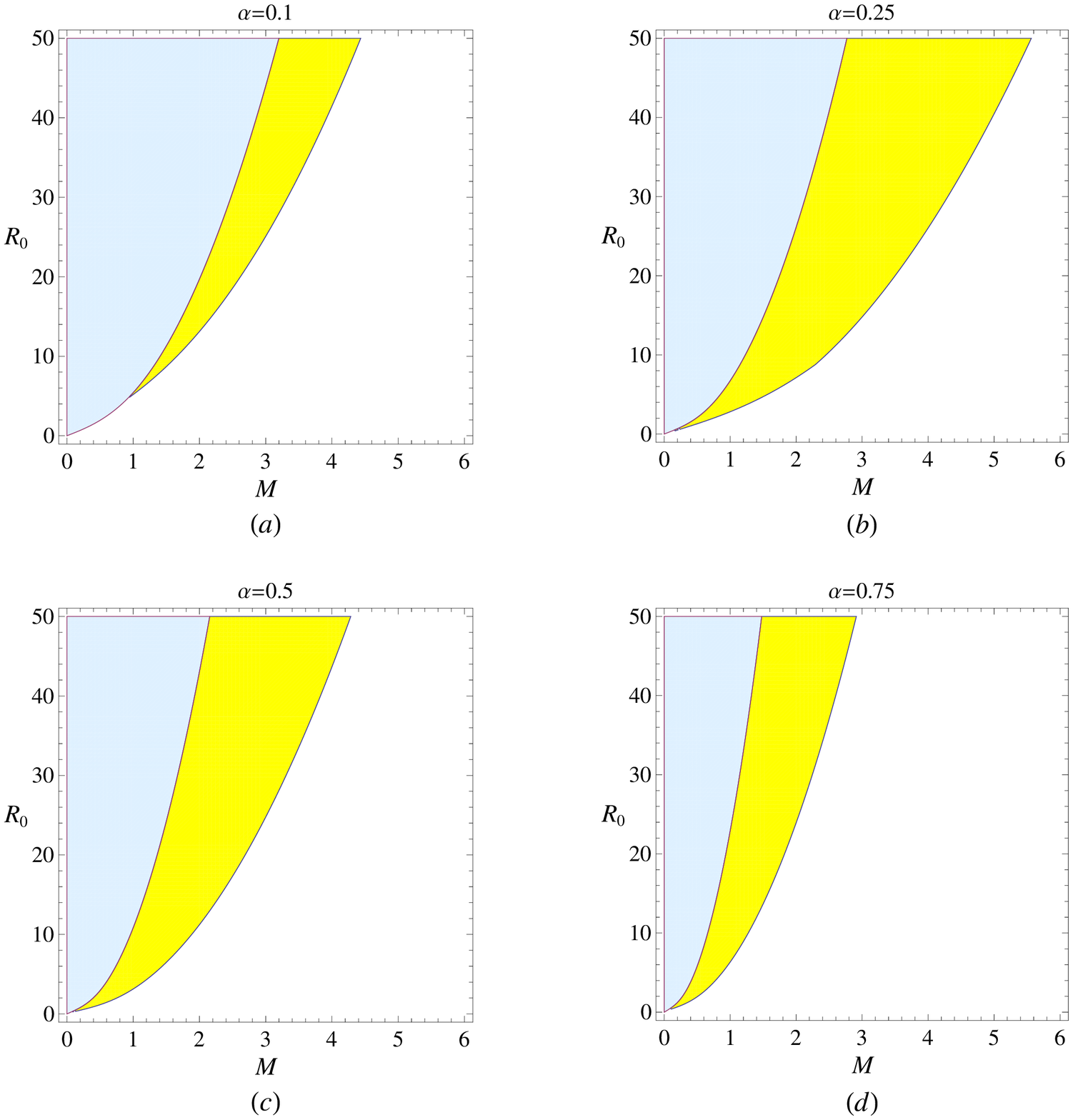}}
\caption{}
\label{fig:1}
\end{figure}

\newpage

\begin{figure}
\resizebox{1\columnwidth}{!}{\includegraphics{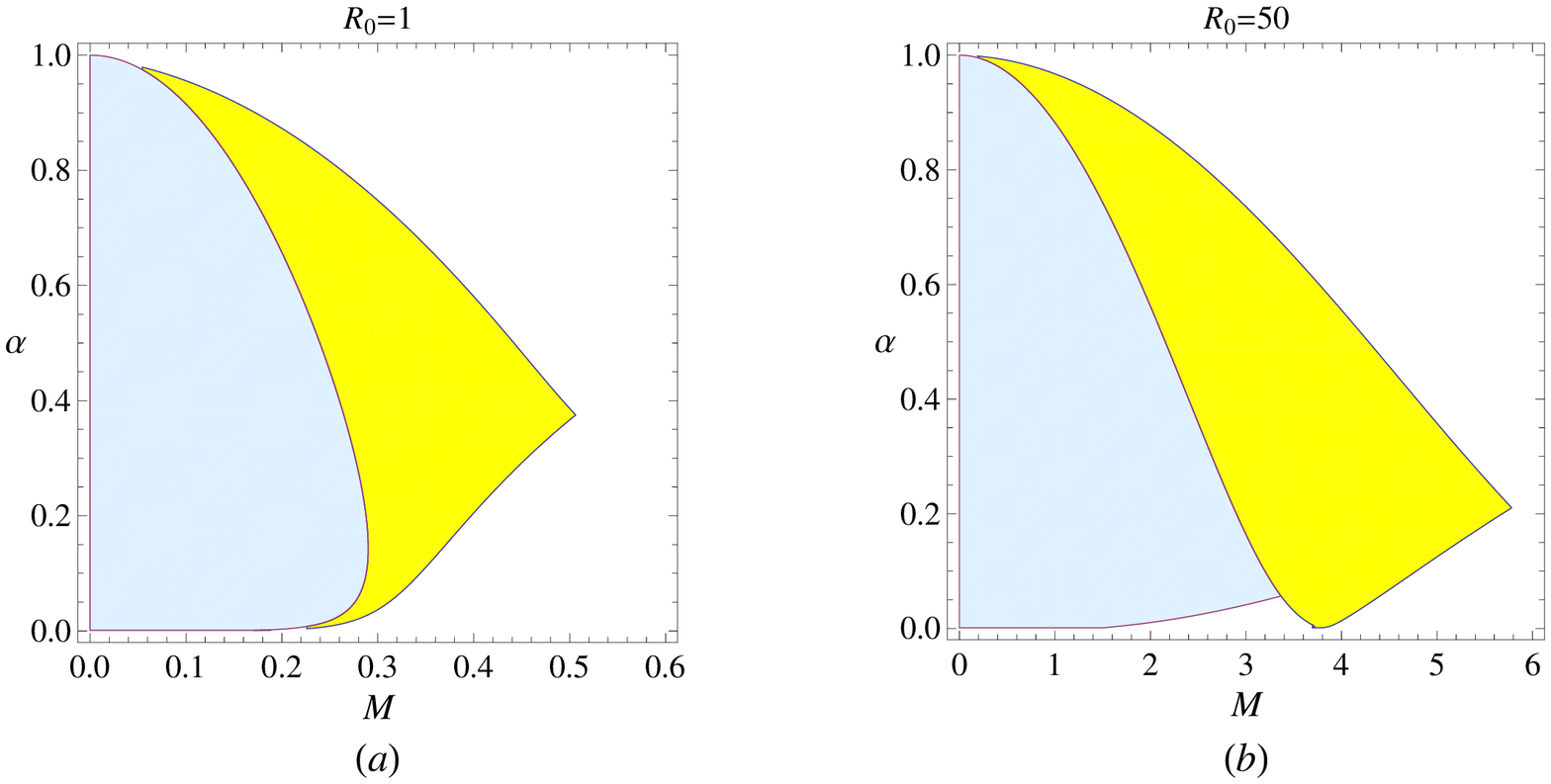}}
\caption{}
\label{fig:2}
\end{figure}

\newpage

\begin{figure}
\resizebox{1\columnwidth}{!}{\includegraphics{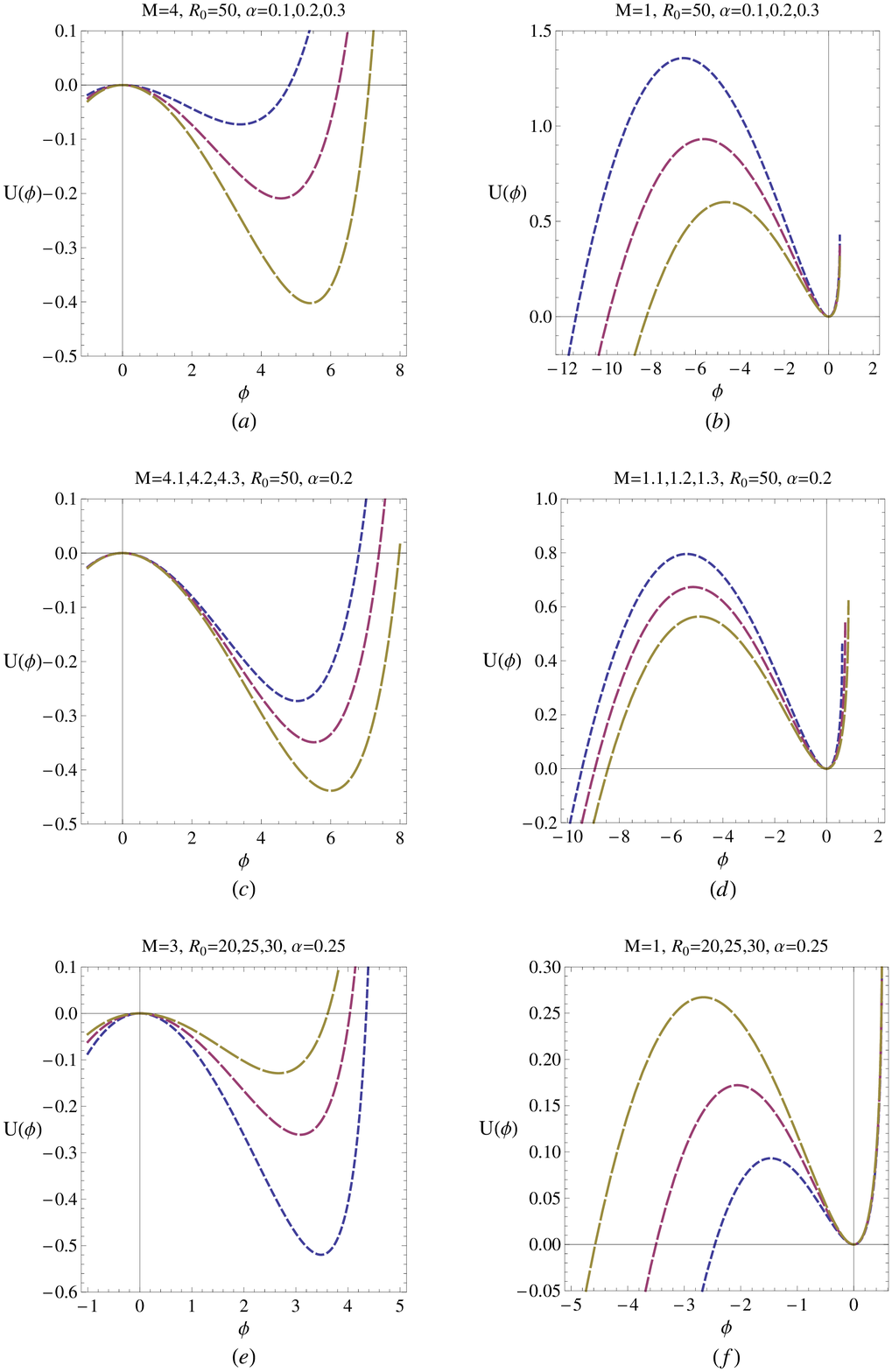}}
\caption{}
\label{fig:3}
\end{figure}

\newpage

\begin{figure}
\resizebox{1\columnwidth}{!}{\includegraphics{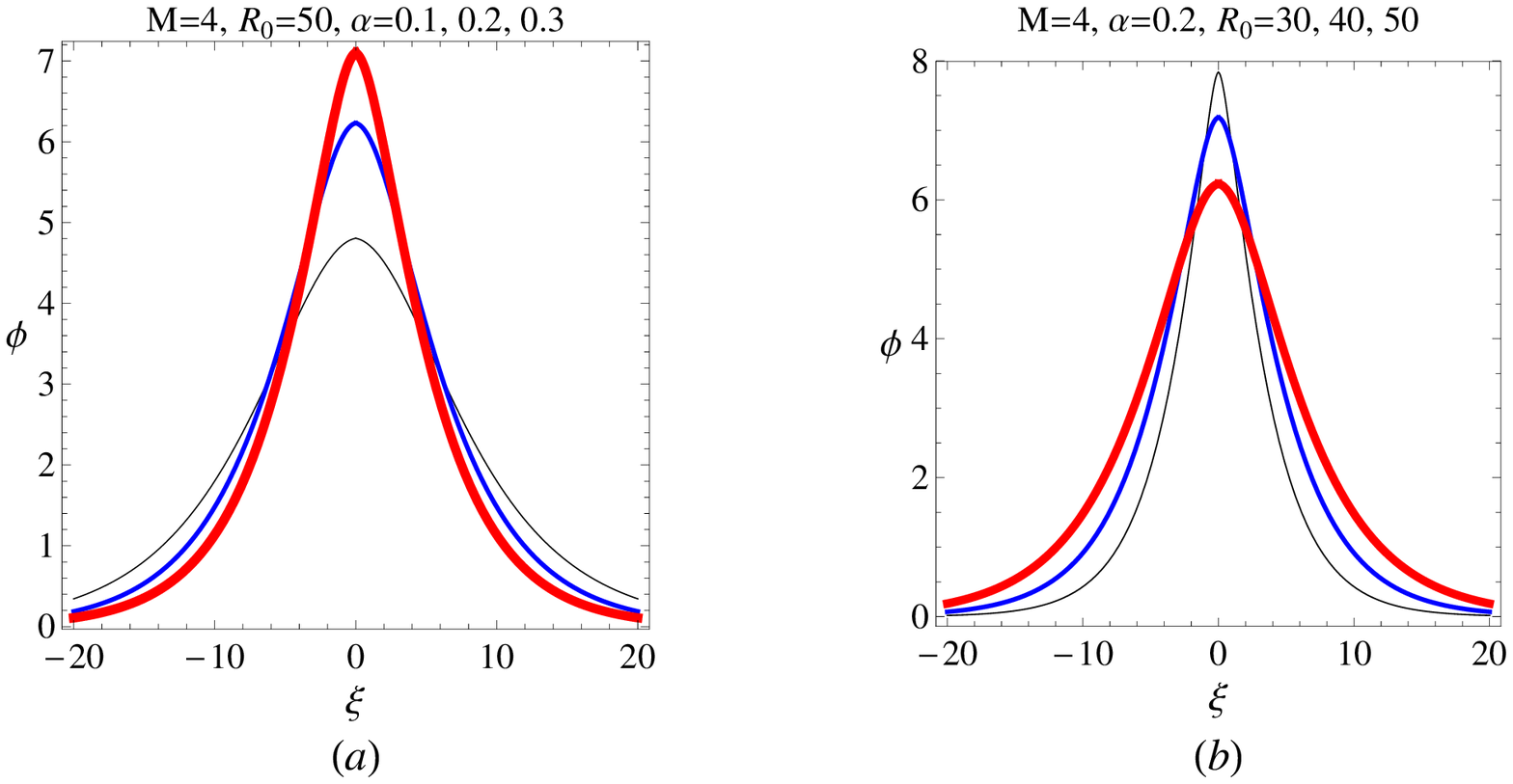}}
\caption{}
\label{fig:4}
\end{figure}

\newpage

\begin{figure}
\resizebox{1\columnwidth}{!}{\includegraphics{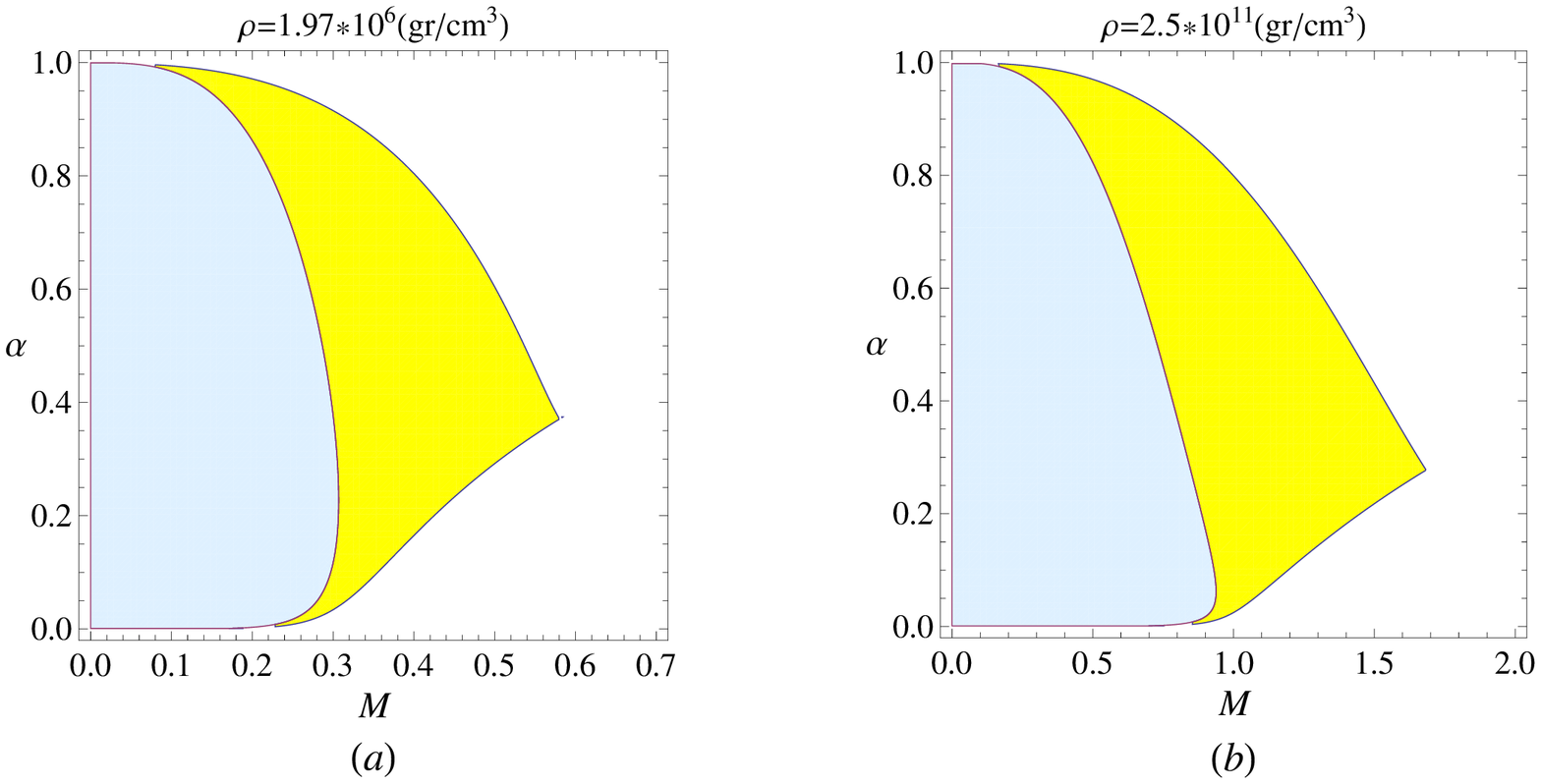}}
\caption{}
\label{fig:5}
\end{figure}

\end{document}